\documentclass[reprint,superscriptaddress,amsmath,amssymb,aps]{revtex4-2}

\usepackage{graphicx}
\usepackage{dcolumn}
\usepackage{bm}

\usepackage{bm}

\usepackage{algorithm}
\usepackage[noend]{algpseudocode} 
\usepackage{amssymb}

\usepackage{tabularx}
\usepackage{graphicx}
\usepackage{dcolumn}
\usepackage{bm}
\usepackage{amsmath}
\usepackage{xcolor} 
\usepackage[pagebackref]{hyperref} 
\hypersetup{
	colorlinks   = true, 
	urlcolor     = blue, 
	linkcolor    = blue, 
	citecolor   = blue 
}
\usepackage{graphicx}
\usepackage{physics}
\usepackage{mathtools}
\usepackage{dsfont}
\usepackage{soul} 
\usepackage{natbib} 
\begin{document}

\preprint{APS/123-QED}

\title{Towards reconstructing quantum structured light on a quantum computer}
\author{Mwezi Koni}
\affiliation{School of Physics, University of the Witwatersrand, Private Bag 3, Wits 2050, South Africa}
\author{Shawal Kassim}
\affiliation{School of Computer Science and Applied Mathematics, University of the Witwatersrand,
Johannesburg, South Africa}
\author{Paola C. Obando }
\affiliation{School of Physics, University of the Witwatersrand, Private Bag 3, Wits 2050, South Africa}
\author{Neelan Gounden}
\affiliation{School of Physics, University of the Witwatersrand, Private Bag 3, Wits 2050, South Africa}

\author{Montaz Ali}
\affiliation{School of Computer Science and Applied Mathematics, University of the Witwatersrand,
Johannesburg, South Africa}
\author{Isaac Nape}
\email{isaac.nape@wits.ac.za}
\affiliation{School of Physics, University of the Witwatersrand, Private Bag 3, Wits 2050, South Africa}


\begin{abstract}
We introduce a variational quantum computing approach for quantum state reconstruction within a discretized logical framework, using experimental measurement data as input. By mapping the reconstruction cost function onto an Ising model, the problem can be solved using a variational eigensolver on present-day quantum hardware identifying dominant logical elements of the density matrix. As a proof of concept, we demonstrate the method on quantum structured light, in particular, entangled photons carrying orbital angular momentum and show that the reconstruction procedure can yield reliable performance even on noisy devices. Our results highlight the potential of variational algorithms as a complementary approach to quantum state tomography, particularly for high-dimensional structured light, where classical approaches can face bottlenecks.
\end{abstract}

\maketitle
\date{\today}

\section{Introduction}
Quantum technologies have become a promising tool for advancements in various fields, which include communications \cite{forbes2024quantum,mafu2013higher,pirandola2020advances}, computing \cite{o2009photonic,nielsen_chuang_2010,ladd2010quantum}, imaging \cite{defienne2024advances} and metrology; promising unprecedented security for communication channels \cite{scarani2009security,renner2008security}, exponential speed up for data processing \cite{shor1999polynomial}, and high resolution in imaging systems \cite{kolobov2000quantum,giovannetti2004quantum}. In computing, certain quantum algorithms promise to solve problems faster than their classical counterparts. Grover's algorithm provides a quadratic speedup for database search while Shor's algorithm yields an exponential speedup for factoring prime numbers \cite{shor1999polynomial,grover1996fast}. These advancements necessitate robust methods for characterising quantum systems to ensure their reliability and performance. Quantum State Tomography (QST) is crucial for this purpose \cite{toninelli2019concepts}. QST allows for the measurement and verification of quantum states (and in some cases as a subroutine for channels and processes) \cite{james2001,agnew2011tomography,poyatos1997complete}. Here a complete set of measurements is performed on the system, and from these the underlying density matrix is reconstructed \cite{toninelli2019concepts}.

While various reconstruction techniques, including Maximum Likehood Estimation (MLE) \cite{hradil1997quantum}; Bayesian methods \cite{buvzek1998reconstruction,jones1991principles} and machine learning \cite{lohani2020machine,neugebauer2020neural} inspired approaches, have been developed to this end, least squares inversion- a technique that solves a system of linear equations relating measured quantities with a density matrix to be constructed \cite{opatrny1997least}, remains one of the popular techniques.

There has been a growing interest in solving linear equations on a quantum computer. Notably, Harrow-Hassidim-Lloyd (HHL) proposed an algorithm that scales logarithmically in the number of unknown parameters, promising quantum advantage for sparse matrices \cite{harrow2009quantum}. The algorithm has recently been implemented in small-scale problems, solving 2-dimensional problems, on various hardware platforms, in superconducting and nuclear magnetic resonance processors \cite{neugebauer2020neural,zheng2017solving}. However, the circuit depth involved renders the algorithm impractical for current noisy hardware with a limited number of qubits for error correction  to be able solve any useful problem.
We make use of current noisy hardware through Variational Quantum Algorithms (VQAs). These are hybrid classical-quantum algorithms, proposed by Peruzzo et al \cite{peruzzo2014}, initially introduced as the Variational Quantum Eigensolver (VQE).
The VQAs make use of shallow parameterized circuits to evaluate a cost function, while a classical optimizer updates its parameters iteratively until convergence towards the solution. Algorithms of this form have been widely applied across quantum chemistry, combinatorial optimization, and machine learning. Demonstrating their versatility as near-term quantum tools \cite{cerezo2021variational,bharti2022noisy,bezuidenhout2024variational,cimini2024variational,kim2024qudit}.
\begin{figure*}[!ht]
		\centering		\includegraphics[width=1\linewidth]{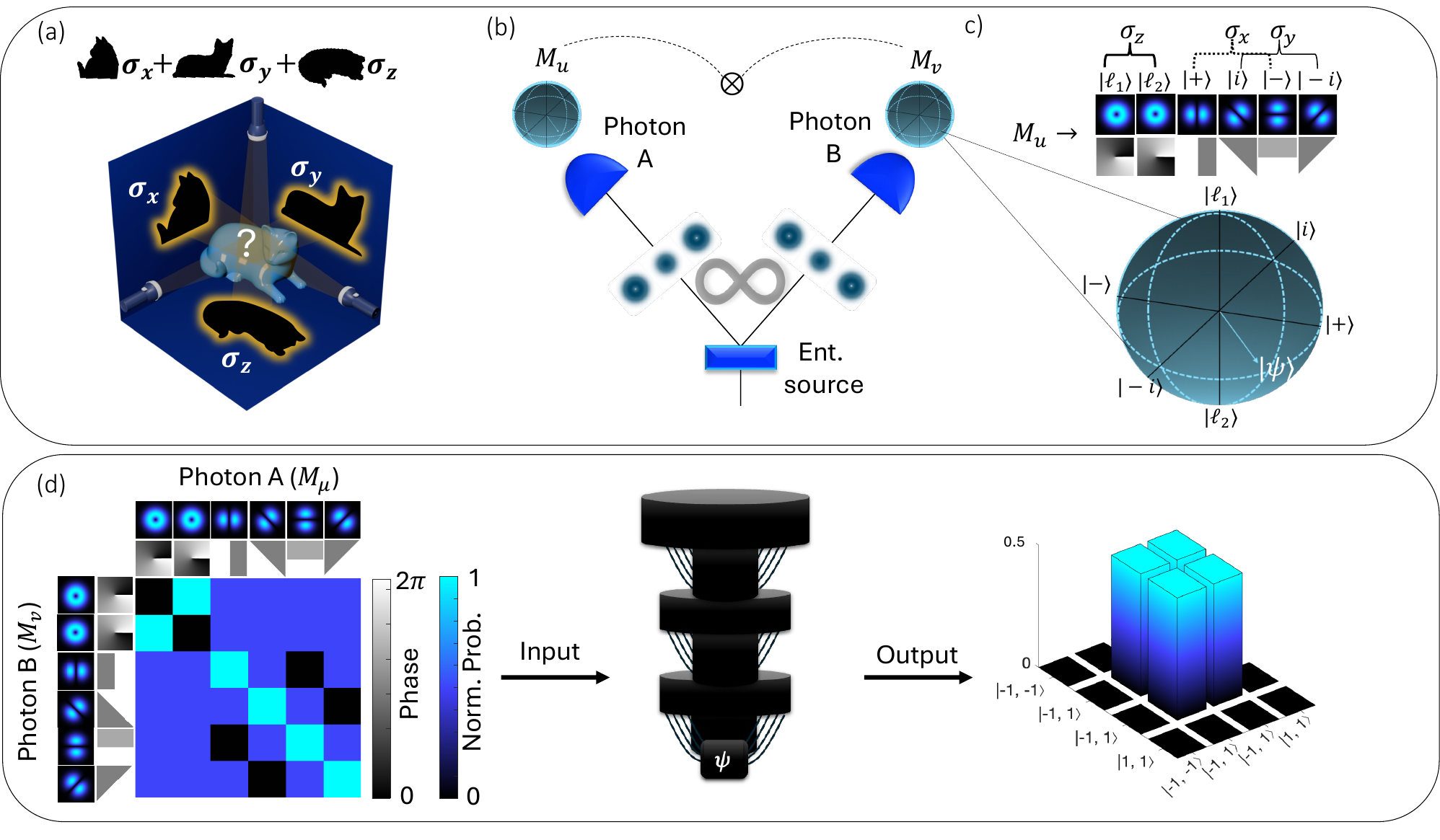}
  \caption[Conceptual illustration of tomography scheme.] {(a) We highlight the conceptual layout of the tomography problem. Similar to reconstructing an image of a 3D object from its projections, in QST we aim to reconstruct the density matrix $\rho$ (our version of the 3D object) from a set of observables. (b) The underlying state we wish to uncover is that of a two-photon state described within the OAM degree of freedom. The two photons here are entangled. Assuming that each photon is defined using a two two-level system of OAM states, each photon is measured using a projection $M^{\pm}_{u,v}$, collapsing each photon on the Bloch sphere. (c) Representation of a two-level system of OAM states on the Bloch sphere, which is equivalent to that for standard qubit states. The states selected on the sphere (and shown in the inset) constitute an overcomplete set of measurements obtained from the eigenvalues of the Pauli operators. (d) Using the joint measurement outcomes $\langle P_k \rangle \equiv \langle M^{\pm}_u \otimes M^{\pm}_{v} \rangle$ for photons A and B, an optimisation routine is implemented and executed on the quantum computer to find the underlying density matrix.}
  \label{fig:conceptfig}
  \end{figure*}

It is natural to ask whether VQAs can also be employed for state reconstruction, especially for high-dimensional states such as structured light fields \cite{nape2023, toninelli2019concepts, forbes2019qhd, he2022towards}, which reside in Hilbert spaces of dimension greater than $d=2$. Such states play a central role in enabling ultrasecure quantum communication channels \cite{wang2020satellite}, high-dimensional quantum computing \cite{lib2024resource}, and quantum imaging \cite{erhard2020advances}. 
In the absence of prior assumptions about the state, the demand of resources for state reconstruction grows exponentially in both the number of required measurements and optimization parameters. This scales as $\mathcal{O} (d^n)$ for $n$ particles, leading to significantly increased measurement times and computational costs.

This work introduces a variational quantum computing methodology for quantum state reconstruction within a restricted logical framework. A set of complete experimental measurements, on an unknown quantum state, are fed as an input from which the algorithm identifies the dominant logical components of the density matrix most consistent with the measured projection.
This is done for $n=2$ particles (photons), each occupying $d=2$ dimensional states. Starting from the least-squares formulation, we derive an explicit algebraic mapping from measurement data to an Ising Hamiltonian and implement a VQE-based reconstruction scheme. We validate our approach by reconstructing structured photons carrying Orbital-Angular-Momentum (OAM). The two photons are entangled and generated via Spontaneous Parametric Down-Conversion (SPDC) where the collected data is in the form of classical joint measurement outcomes (photon coincidence counts). The state reconstruction procedure is then performed on a superconducting qubit-based quantum computer, where we examine three architectures, specifically, based on a single depth rotation gate $R_y$, depth three $R_y$ rotation gates, and a universal single qubit rotation \{$R_y$, $R_z$, $R_y$\} gates of depth three ansatz families. Our results serve as a proof of concept, showing that variational algorithms can be employed for state reconstruction on near-term devices. Although the present demonstration does not show a performance advantage over classical reconstructions at this system size, the algebraic mapping and VQE pipeline presented here establish a flexible platform for future work on scalable encodings, noise mitigation, and hybrid strategies that may render variational tomography competitive for larger quantum systems.
 
\section{Theory}\label{sec:hdst}
\begin{figure*}[ht!]
		\centering
        \includegraphics[width=1\linewidth]{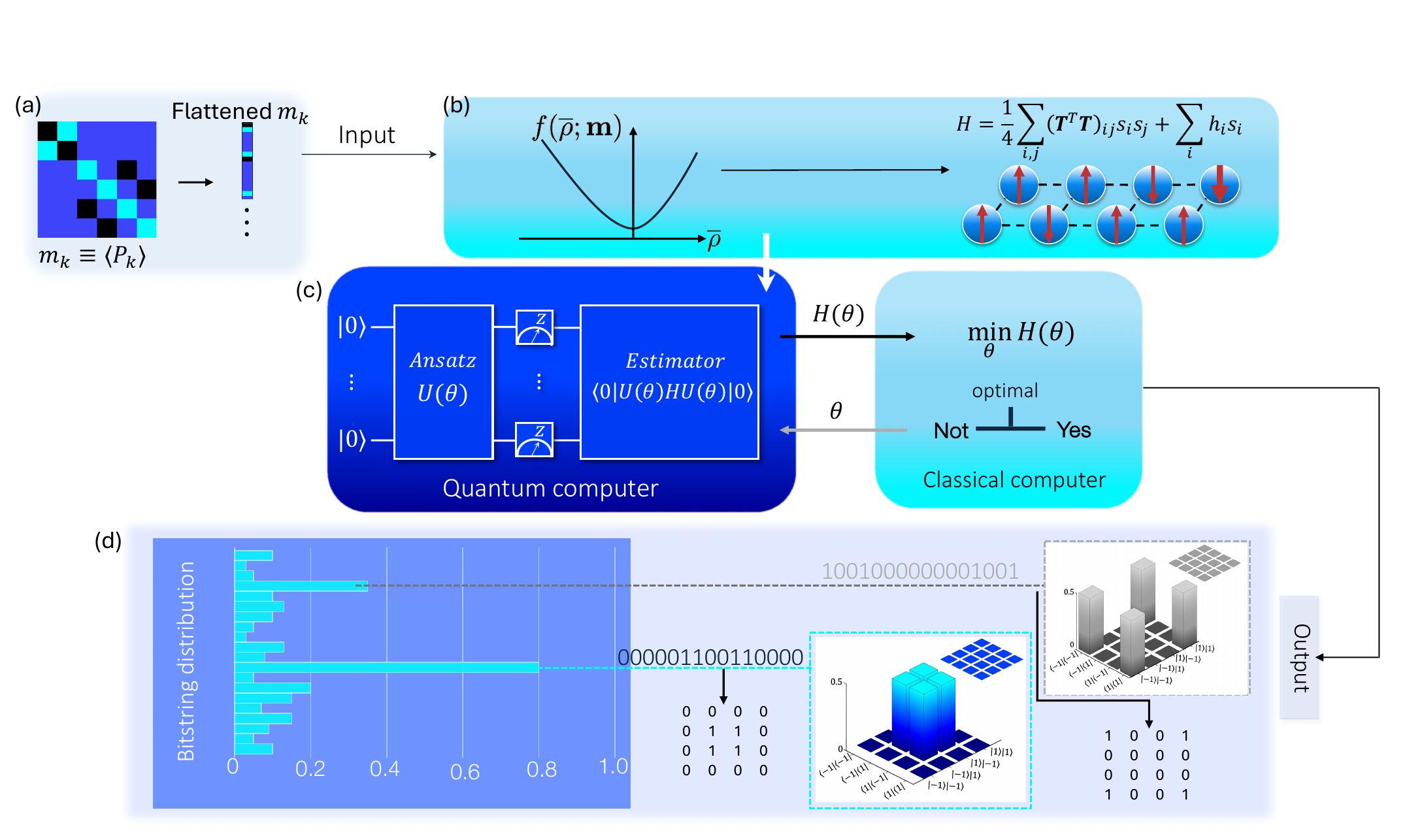}
  \caption[VQE Algorithm Workflow]{VQE Algorithm Workflow: (a) The input variables are transformed into spin variables, converting the least-squares problem into an Ising spin model. (b) The cost function is expressed as a Hamiltonian in terms of spin observables. (c) A quantum computer evaluates the Hamiltonian's expectation value based on ansatz circuit parameters, which are iteratively optimized by a classical computer. (d) Once the expectation value is minimized, the optimized circuit parameters are used to sample the bitstring distribution through measurements in the computational basis. The density matrix corresponding to bitstrings with the highest counts is used to obtain the density matrix representation of the best solution.}
  \label{fig:workflow}
	\end{figure*}
    
\subsection{State tomography of spatial modes}
We begin by introducing the fundamental concepts of Quantum State Tomography (QST) for photonic states carrying Orbital Angular Momentum (OAM).
 After the fundamentals are established, we demonstrate how the state reconstruction procedure can be implemented.

A process-level conceptual
representation of QST is illustrated in Fig.~\ref{fig:conceptfig}(a).
Here, an unknown state is probed using observables, visualised as shadows of the unknown object (representing the state). The measurements form a Positive Operator-Valued Measure (POVM), capable of sampling the essential features of the state. 
In the context of qubits, the projections are observables of the Pauli matrices, $\sigma_{x}, \sigma_{y}, \sigma_{z}$, with each matrix having a spectral decomposition, $\sigma_j = M^{+}_{j} - M^{-}_{j}$ ($j\in \{x, y, z\}$), where $M^{\pm}_{j}$ are the corresponding eigenvectors of the j$^{th}$ Pauli matrix. In practice, one measures the outcomes of these projectors in the experiment. 
For an object in 3-D, this is analagous to casting
shadows onto multiple planes.
In contrast, for a quantum state, it would correspond to constructing a set of projective measurements that sample components of the state on the Bloch sphere, which contain essential information about the density operator $\hat{\rho}$ of the underlying quantum state.
Our work
aims to apply this procedure to a two-photon state with each photon
occupying a two-dimensional space.

We focus on entangled photons encoded in the OAM basis as illustrated in Fig.~\ref{fig:conceptfig} (b). Assuming the photons are anti-correlated in OAM, we can describe the state as a linear combination of two-dimensional entangled Bell states following
\begin{equation}
   \ket{\psi}_{AB} = \sum_{\ell=0}^{L} c_{|\ell|} \left( \ket{\ell}_A \ket{-\ell}_B + \ket{-\ell}_A \ket{\ell}_B \right),
\end{equation}
where $c_{|\ell|}$ are coefficients weighting each Bell state, $\ket{\ell}\ket{-\ell} + \ket{-\ell}\ket{\ell}$. The OAM basis modes, $\ket{\pm\ell} \propto \int A(r) \exp(\pm i \ell \phi) \ket{\phi}\ket{r} d^2r$, corresponding to eigenstates of the OAM operator, are described by cylindrically symmetric wavefunctions, $A(r)\exp( \pm i \ell \phi)$, in polar coordinates, $(r, \phi)$. Importantly, the OAM basis states carry an azimuth-dependent ($\phi$) phase function, $\exp( \pm i\ell \phi)$, corresponding to a photon field having an OAM of $\pm \ell \hbar$ per photon. Accordingly, each photon inhabits an infinite Hilbert space, $\mathcal{H}_{A(B)}=\mathrm{span}\{\ket{\ell}_{A(B)}:\ell\in\mathbb{Z}\}$. For the purpose of this demonstration, we restrict the subspace spanned to two dimensions so that two elements span the state space $\{\ket{\ell},\ket{-\ell}\}$. Fig. \ref{fig:conceptfig} (c) shows the OAM Bloch sphere \cite{padgett1999poincare} corresponding to a preselected subspace of $\ell = 1$ for a single photon. The states shown in the figure correspond to the projection states, i.e., 
\begin{align}
    M_{z}^{\pm } &\rightarrow \ket{\pm\ell}, \nonumber \\ 
     M_{x}^{\pm } &\rightarrow \ket{\pm} = \left( \ket{\ell} \pm \ket{-\ell} \right) /\sqrt{2},  \nonumber \\ 
          M_{y}^{\pm }, &\rightarrow \ket{\pm i} =  \left( \ket{\ell} \pm i \ket{-\ell}  \right) /\sqrt{2}. \nonumber \\ 
\end{align}
Where the logical/computational basis states, which are eigenstates of $\sigma_z$ are on the poles of the Bloch sphere, whereas the eigenstates of $\sigma_{x, y}$ are all on the equator.
Represented in Fig. \ref{fig:conceptfig} (c), are the six overcomplete measurements that are needed in QST.
We note that at least four measurements, forming a POVM are sufficient and complete. For two photons, we can construct the set of overcomplete measurements $P_k=M^{\pm}_u\otimes M^{\pm}_v$ from tensor products of these measurements, yielding a total of 36 overcomplete measurements as shown in the first panel of Fig. \ref{fig:conceptfig} (d). 
The measurements
are mapped onto the quantum computer to find the underlying quantum
state, represented as a density matrix.
This is shown conceptually in the last two panels of the Fig. \ref{fig:conceptfig} (d). 
The reconstruction of the density matrix is accomplished through
minimising an appropriate cost function, discussed in the next section.

\subsection{From quadratic cost function to the energy Hamiltonian }
Given the measurement outcomes $m_k = \langle P_k \rangle$, we now describe how Quantum State Tomography (QST) can be formulated as a Quadratic Unconstrained Binary Optimisation (QUBO) problem. 
In this framework, the measurement probabilities serve as inputs, as illustrated in Fig.~\ref{fig:workflow}(a), and the optimisation seeks the underlying quantum state $\rho$ by minimising a quadratic cost function. This cost is then re-expressed as an energy function in terms of an Ising spin Hamiltonian, shown in Fig.~\ref{fig:workflow}(b). Such a mapping allows us to harness the quantum computer as a solver, effectively searching (illustrated in Fig.~\ref{fig:workflow}(c)) for the state that best minimises the energy function as depicted in Fig.~\ref{fig:workflow}(d)).

Firstly, notice that the measurement outcomes obey Born's rule following
\begin{equation}
    m_k = \mathrm{tr}(P_k \hat{\rho}),
\end{equation}
which, via the Hilbert–Schmidt inner product, can equivalently be written as a dot product between vectorized operators,
\begin{equation}
    m_k = \langle P_k,\hat{\rho}\rangle_{\mathrm{HS}} = \overline{P}_k\cdot\overline{\rho},
\end{equation}
where $\overline{P}_k=\operatorname{vec}(P)$ and $\overline{\rho}=\operatorname{vec}(\hat{\rho})$, are vector (flattened) representations of the measurement projectors and density matrix, respectively. Collecting all $K$ outcomes into $\mathbf{m}\in\mathbb{R}^{K\times1}$ and stacking $\overline{P}_k$ as the rows of a measurement matrix defines the linear model
\begin{equation}
    \label{eq:linear_model}
    \mathbf{m} = T\,\overline{\rho}, \qquad T\in\mathbb{C}^{K\times N},\; N=d^2,
\end{equation}
where $T$ contains stacked rows of the measurement matrices, i.e.
having the form 
\begin{equation}
T = \begin{bmatrix}
- \bar{P}_0^{\dagger} - \\
.\\ 
.\\ 
- \bar{P}_{K}^{\dagger} -
\end{bmatrix}.
\end{equation}
Reconstruction by least squares minimises the quadratic cost
\begin{equation}
    \label{eq:lsq}
    f(\overline{\rho}; \mathbf{m}) = \|\mathbf{m}-T\overline{\rho}\|_2^2
    = \overline{\rho}^\dagger Q\,\overline{\rho} - 2\,\mathrm{Re}\!\big(\mathbf{m}^\dagger T \overline{\rho}\big) + \mathbf{m}^\dagger\mathbf{m},
\end{equation}
with $Q\equiv T^\dagger T$. 
 Therefore, given the measurement matrix $T$ and the observed frequencies $\mathbf{m}$, minimizing the least-squares objective in Eq. \eqref{eq:lsq} (subject to the physicality constraints $\hat{\rho}> 0$ and $\mathrm{tr}\,\hat{\rho}=1$) yields an estimate of the density operator. 

Next, the goal is to map the quadratic cost function onto an Ising Hamiltonian as depicted in Fig. \ref{fig:workflow} (b). To do this, we follow the approach outlined in \cite{lucas2014ising}. The Hamiltonian can be decomposed using tensor products of Pauli matrices $\sigma_{\alpha_j}$, following
\begin{equation}\label{eq:H}
    H=\sum_\alpha h_\alpha \otimes^{N}_{j=1}\sigma_{\alpha_j},
\end{equation}
where $\sigma_\alpha$ are the Pauli operators which include the identity, $\otimes^{N}_{j=1}$ represents the tensor product over $N$ qubits with $j$ indexing each qubit, whereas $h_\alpha$ are the coefficients that determine the Hamiltonian for each contributing term. In this work, we restrict the composition of each tensor product term to sequences of $\sigma_z$ matrices and the identity $\mathbb{I}_2$ in which we adopt the notation, $Z_j$ denoting the single-qubit (or single body) Pauli-$Z$ operator acting on logical qubit $j$ and identity on the remaining qubits, i.e $Z_j \;= (\otimes^{j-1}_{k} \mathbb{I}_2) \otimes \sigma_z \otimes (\otimes_{k}^{j+2} \mathbb{I}_2)$. Therefore, a two-body Pauli string $Z_j Z_k$ denotes $Z$ acting on qubits $j$ and $k$ and identity elsewhere.


With this convention established, we convert our cost function in Eq.~\eqref{eq:lsq}, $f(\bar{\rho}, \mathbf{m})$, onto the appropriate Hamiltonian, by first mapping the entries of the flattened density matrix, $p_j$, onto the $\sigma_z$ expectation values through an affine transformation following
\begin{align}
    \label{eq:affine_encoding}
    p_j &\longleftrightarrow \frac{1 - \langle Z_j\rangle}{2}, \qquad j=1,\dots,N_{\mathrm{enc}}, \nonumber\\
\end{align}
so that $\langle Z_j\rangle\in[-1,1]$ corresponds to $p_j\in[0,1]$.
To make the encoding explicit, let $N$ denote the number of real scalar variables that we place on the quantum device. In the present work, we adopt a one-to-one (basis) encoding and therefore set the number of qubits to $N_{\mathrm{enc}}$ (equivalent to the number of elements in the density matrix), in the variational circuit; consequently, the index $j$ below runs over $j=1,\dots,N$. Under this convention, each encoded scalar $p_j$ is associated with a single qubit via the affine transformation in Eq. (\ref{eq:affine_encoding}) so that we can map the components of the density matrix onto operators, i.e. $p_j \mapsto  \frac{1 -  Z_j}{2}$. Under this mapping, the pairwise products satisfy
\begin{equation}
    \label{eq:pairwise}
    p_j p_k \mapsto \frac{1}{4}\big(1 - Z_j - Z_k + Z_j Z_k\big).
\end{equation} Substituting these mappings into the quadratic form (Eq. \eqref{eq:lsq}) expressed in the variables $p_j$ and collecting terms, generates an operator $\mathcal{H}$ that is diagonal in the computational basis and whose expectation value reproduces the least-squares cost up to an additive constant as desired. Writing $Q_{jk}$ for the entries of $Q$ and $t_j$ for entries of $t$, the Hamiltonian takes the Ising (Pauli-$Z$) form
\begin{equation}
    \label{eq:ising}
    \mathcal{H} = \sum_{j<k} J_{jk}\,Z_j Z_k + \sum_j h_j\,Z_j + \mathrm{offset},
\end{equation} with coefficients given directly by $Q$ and $t$:
\begin{align}
    \label{eq:J}
    J_{jk} &= \tfrac{1}{4}\, \Re(Q_{jk}), \qquad (j\neq k),\\[6pt]
    \label{eq:h}
    h_j    &= -\tfrac{1}{2}\sum_{k} \Re(Q_{jk}) + \mathrm{Re}(t_j),\\[6pt]
    \label{eq:offset}
    \mathrm{offset} &= \tfrac{1}{4}\sum_{j,k} Re(Q_{jk}) + \mathbf{m}^\dagger\mathbf{m} - \sum_j \mathrm{Re}(t_j).
\end{align}
where Re$(\cdot)$ represents real parts. If complex flattened entries are encoded, real and imaginary parts should be handled as separate encoded scalars; the expressions above assume encoding of real-valued quantities. The variational objective is the Hamiltonian expectation on a parameterized pure state $|\psi(\boldsymbol{\theta})\rangle$:
\begin{equation}
    \label{eq:variational}
    E(\boldsymbol{\theta}) = \langle \psi(\boldsymbol{\theta}) | \mathcal{H} | \psi(\boldsymbol{\theta}) \rangle.
\end{equation} Minimization is performed in a hybrid scheme as illustrated in Fig.~\ref{fig:workflow} (c): a parameterized quantum circuit prepares the state $|\psi(\boldsymbol{\theta})\rangle$ by applying a paramertrised unitary $U(\boldsymbol{\theta})$ to the $N$ qubits initialsed as $|0\rangle^{\otimes N}$. In our case the unitary is built from the three ansatz families composed of single depth rotation gate $R_y(\boldsymbol{\theta}_1)$, depth three $R_y$ gates $R_y(\boldsymbol{\theta}_1),R_y(\boldsymbol{\theta}_2),R_y(\boldsymbol{\theta}_3) $ rotation gates and a universal single-qubit rotation gate set $\{R_z(\boldsymbol{\theta}_1), R_y(\boldsymbol{\theta}_2), R_z(\boldsymbol{\theta}_3)\}$ of depth three. The quantum processor (or simulator) provides estimates of the expectation values $\langle Z_j\rangle$ and $\langle Z_j Z_k\rangle$, and the classical optimizer updates $\boldsymbol{\theta}$ to minimise the energy/cost, $E(\boldsymbol{\theta})$. After convergence, sampling the optimized circuit in the computational basis yields a distribution over possible basis states in the logical basis. The states are interpreted as bitstrings from which we extract the one with the highest counts, and subsequently reshape it to recover estimates of the encoded binary scalar variables, thereby reconstructing the desired density matrix components.

As an illustrative example, consider a variational register consisting of $N=4$ qubits. This register spans $2^4 = 16$ possible computational basis states $\ket{0000}, \ket{0001}, \ldots, \ket{1110}, \ket{1111}$ each corresponding to a distinct binary encoding of the vectorized density-matrix parameters. In this example, each bitstring represents a candidate discretized approximation to a  $2 \times 2$ density matrix (i.e. a single-qubit physical system). If the bitstring $\ket{1111}$ inimizes the Ising energy, it corresponds, upon reshaping, to a matrix proportional to $
\rho \propto 
\begin{pmatrix} 
1 & 1 \\ 
1 & 1 
\end{pmatrix}.$ More generally, higher-dimensional density matrices require a larger variational register. For instance, reconstructing a
$4 \times 4$ density matrix (corresponding to a two-qubit physical system) involves encoding 16 independent matrix elements, which in the present binary scheme leads to a variational space of  $2^{16}$ possible bitstring configurations.  a $4 \times 4$ density matrix Figure 2(d) illustrates this latter case, showing a reconstructed  a $4 \times 4$ density matrix obtained from the probability-weighted distribution over these  $2^{16}$ configurations in the final stage of the workflow.
\section{Experiment}
\subsection{Data acquisition}
\begin{figure}[h!]
	\centering
	\includegraphics[width=1\linewidth]{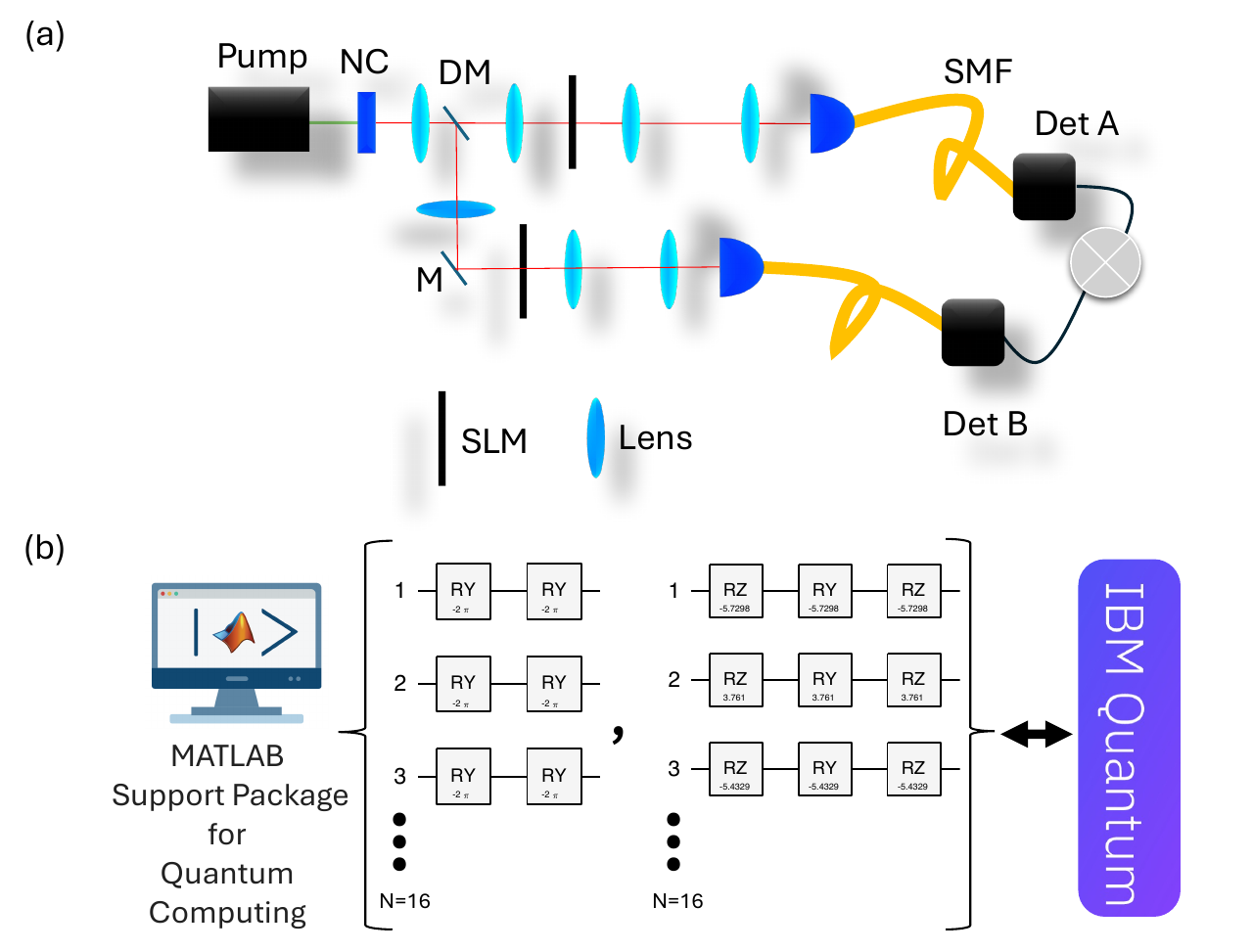}
	\caption{(a) Setup for extracting measurement (\textbf{m}) data using non-degenerate SPDC. A continuous-wave pump laser passes through a Nonlinear Crystal (NC) to generate signal and idler photons at different wavelengths. A Dichroic Mirror (DM) separates the two wavelengths, which are then imaged to Spatial Light Modulators (SLMs). The photons are subsequently imaged to the interface of Single-Mode Fibers (SMF), thereafter coupled and detected by avalanche photodiodes (Det A and Det B) for coincidence counting. (b) Optimisation using the quantum computer. Quantum circuit ansatz prepared using the MATLAB Support Package for Quantum Computing. Circuits of variable depth were generated with either single-qubit $R_Y(\cdot)$ rotations or $R_Z(\cdot)R_Y(\cdot)R_Z(\cdot)$ blocks with adaptable rotation angles. The circuits were executed on IBM Quantum hardware via Qiskit Runtime primitives, using the Estimator primitive for energy minimization and the Sampler primitive for final state extraction.}
	\label{fig:setup}
\end{figure}

\begin{figure*}[t!]\includegraphics[width=1\linewidth]{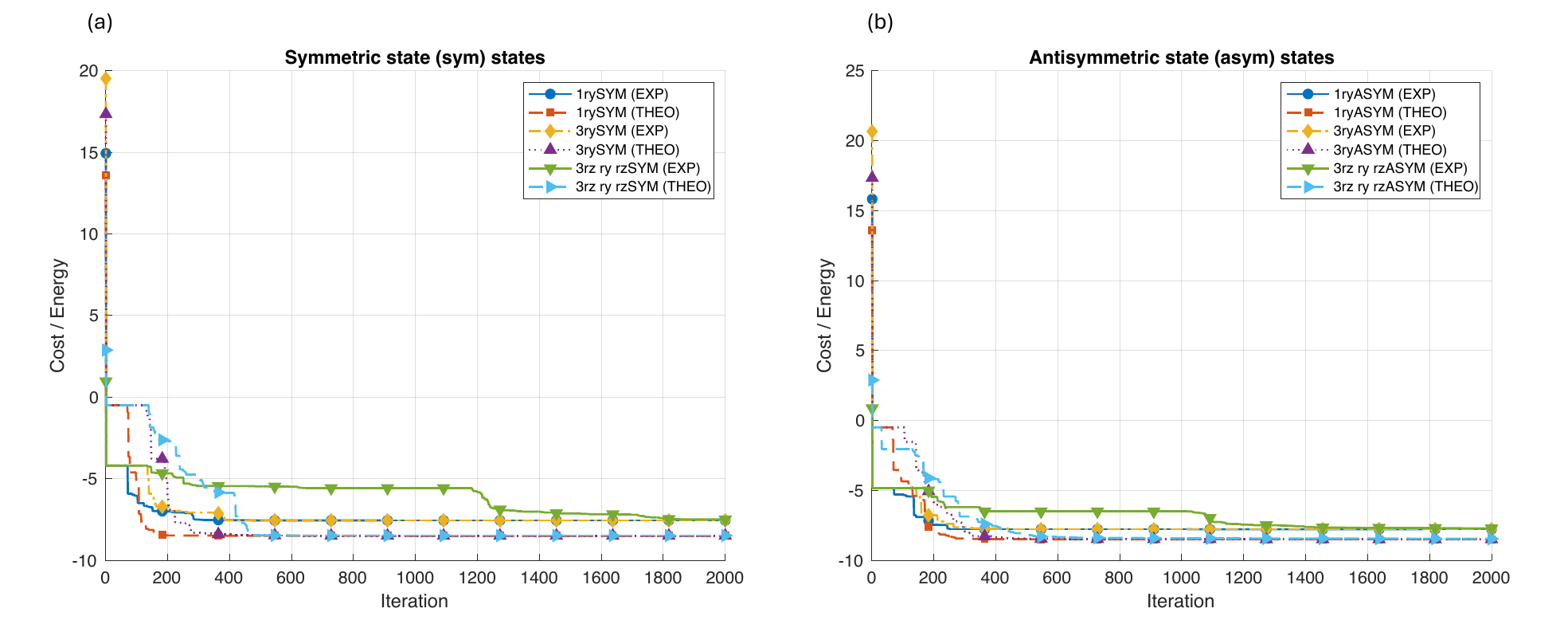}
\caption{Implementation of the VQE-based state-reconstruction scheme. Energy expectation values evaluated at each iteration for anti-correlated (a) and correlated (b) OAM-entangled target states, shown for the different tested circuit architectures. In the legend, the first numeral indicates the number of rotation layers (circuit depth), while the labels “rySYM” and “rz ryASYM” specify the employed gate set and whether symmetric or antisymmetric states are used}
   \label{fig:costf}
\end{figure*}

The measurement outcomes used in this work were collected from a non-degenerate Spontaneous Parametric Down-Conversion (SPDC) experiment, as shown in Fig. \ref{fig:setup} (a). To prepare the entangled state, a continuous-wave diode-pumped laser operating at a wavelength $\lambda_p = 532$ nm was incident on a 5 mm Type-0 PPKTP crystal, which was heated to 60$^\circ$C. Under these conditions, SPDC produced frequency non-degenerate photon pairs at wavelengths $\lambda_A = 1550$ nm (signal) and $\lambda_B = 810$ nm (idler). The photon pairs, emitted collinearly, were separated using a DM that directed the shorter-wavelength photons into one arm and the longer-wavelength photons into another. When both arms contain the same number of mirror reflections, the source produces the anti-correlated OAM entangled stated state, $\ket{\ell}\ket{-\ell}+\ket{-\ell}\ket{\ell}$,as required by orbital angular momentum conservation in SPDC. To generate a correlated OAM state, an additional mirror was introduced in one arm of the setup. This extra reflection implements a mode inversion $\ket{\ell} \rightarrow\ket{-\ell}$ for one photon, thereby transforming the initial anti-correlated state into the correlated basis state $\ket{\ell}\ket{\ell}+\ket{-\ell}\ket{-\ell}$. Projective measurements were carried out independently using the SLM to encode the transmission functions corresponding to the wavefunctions $\exp( \pm i \ell \phi)$, $\exp(  i \ell \phi) + i \exp(-  i \ell\phi)$, $\exp(i \ell \phi) -  \exp(- i \ell\phi)$, and $\exp( i \ell\phi) + i \exp(-i\ell\phi)$ with $\ell = 1$ described in the azimuthal spatial coordinate ($\phi$) of each photon. These transmission functions were encoded on each arm using phase-only encoding \cite{pinnell2020modal}. Each of these masks was loaded on the SLM sequentially. Each SLM modulated the photon with the corresponding pattern, with the modulated photon coupled to a single-mode fiber that only allows the fundamental Gaussian mode to propagate. The SLM and SMF form a projective measurement for spatial patterns of photons \cite{pinnell2020modal, nape2023, roux2014projective}. Subsequently, the fibers were connected to Avalanche Photo-Detectors (APDs) for photon counting. A time-correlated counting module registered coincidences within a 3 ns window between the detectors. Accordingly, a total of 36 measurements were obtained from the systems, which contained coincidence count measurements. These measurements were then normalized and used in the variational algorithm below.

\subsection{Variational Quantum Eigensolver (VQE) implementation}
To validate our VQE-based tomography scheme, we  utilized two complementary data sets: (i) ideal state-vector simulations of a two-qubit Bell state, and (ii) experimental coincidence counts from our SPDC OAM-entangled photon source. Measurement outcomes were assembled into the linear model $\mathbf{m}=T\,\operatorname{vec}(\hat{\rho})$; from which we form $Q = T^\dagger T,\qquad t = T^\dagger \mathbf{m}$,
and map the real part of the quadratic objective to a Pauli-$Z$ Ising Hamiltonian with coefficients $J_{ij} = \tfrac{1}{4}\,\mathrm{Re}(Q_{ij})$, $h_i = -\tfrac{1}{2}\sum_j \mathrm{Re}(Q_{ij}) + \mathrm{Re}(t_i)$ as was derived in the theoretical section. 
\begin{figure*}[ht!]
\centering
\includegraphics[width=1\linewidth]{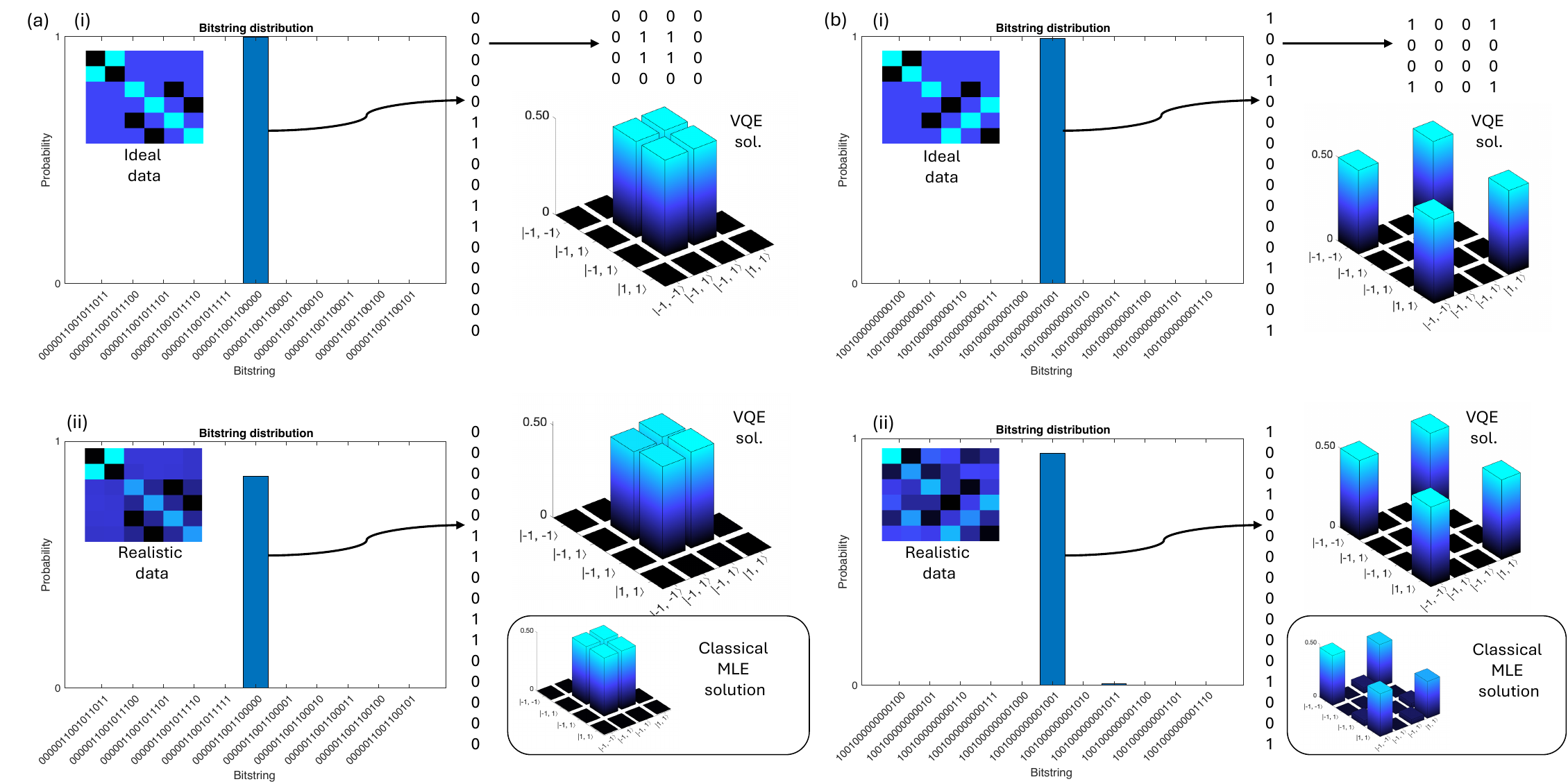}
	\caption{{Sampled bitstring statistics and reconstructed density matrices using VQE.}
		Bar plots show sampled bitstring frequencies for each measurement input (with inset panels displaying the corresponding measurement configurations). (a)(i–ii) Bitstring distributions from simulated and experimental measurements for the antisymmetric state. (b)(i–ii) Bitstring distributions for the symmetric state, with the bitstring having the highest probability mapped onto the corresponding density matrix. In each panel, the reconstructed density matrix is obtained from the weighted sums of the reshaped bitstrings. The corresponding solutions obtained from the quadratic optimisation using the conventional MLE approach are shown for panel (ii),while those in panel (i) match the ideal case.}
	\label{fig:res}
\end{figure*}

Small numerical imaginary residues were discarded; all reported Hamiltonian coefficients are real. The schematic summarising the Ansatz preparation and circuit execution are shown in Fig. \ref{fig:setup} (b), which required interfacing Matlab's quantum computing package with IBM's runtime environment \cite{matlabqc}. Expectation values were evaluated by applying Pauli strings directly to the state vector. For the variational Ansatz, we tested single-layer $R_y$ rotations on each qubit, deeper stacks of $R_y$ layers, and a three-parameter Euler block $\{R_z,R_y,R_z\}$ per qubit. No entangling gates were required, since the Ising Hamiltonian model that we implemented is diagonal in the computational basis. Expectation values $\langle Z_i\rangle$ and $\langle Z_i Z_j\rangle$ were obtained either by direct application of Pauli strings to the statevector (simulation) or by empirical averages of $\pm1$ outcomes from measured bitstrings. In the latter case, the computational basis results were mapped as $|0\rangle \mapsto +1$ and $|1\rangle \mapsto -1$, with the Qiskit runtime primitive Estimator providing the expectation value measurements and once the optimal parameters are found the Sampler provides probability distributions.

For optimisation, we used MATLAB's surrogate optimizer from the Global Optimisation toolbox.
Below we outline the algorithmic framework used to find the optimal parameters that minimize the expectation value of the Hamiltonian. This pseudocode details the iterative process of preparing the quantum circuit with the current set of parameters, measuring the expectation value, extracting the resulting bitstrings, and updating the parameters accordingly until convergence is achieved. By following this structured approach, we efficiently approximated the ground state of the quantum system, obtaining the necessary binary solutions that indicate the presence or absence of components in the density matrix.
\begin{algorithm}[H]
\caption{Variational Quantum Eigensolver (VQE) Algorithm}
\begin{algorithmic}[1]
    \State \textbf{Input:} Initial parameters $\theta$
    \While{not converged}
        \State Prepare quantum circuit with parameters $\theta$
        \State Measure the expectation value $\langle H \rangle$
        \State Update parameters \(\theta\) to minimize $\langle H \rangle$
    \EndWhile
    \State \textbf{Output:} Optimal parameters \(\theta*\) corresponding to the minimal expectation value and extracted bitstrings.
\end{algorithmic}
\end{algorithm}


\section{Results}
To show that we can effectively minimise the energy Hamiltonian (equivalently the cost function) we first present the convergence traces in Fig.~\ref{fig:costf}. Firstly, in Fig. \ref{fig:costf} (a) we show plots of the energy evaluations per iteration for anti-correlated (i), $\ket{-1}\ket{1} +\ket{1}\ket{-1}$, and correlated (ii), $\ket{1}\ket{1} +\ket{-1}\ket{-1}$ OAM-entangled targets across the tested single-qubit rotation architectures. We observe reliable convergence for circuits built from a single-layer $\{\mathrm{R_y}\}$ rotation block at shallow depth and for the nested $\{\mathrm{R_z},\mathrm{R_y},\mathrm{R_z}\}$ block at larger depth corresponding to 1 layer and 3 layers. The cost function evaluations indicate that the least-expensive architecture (single $\{\mathrm{R_y}\}$ layer) converges faster than deeper rotation blocks, and that runs driven by experimentally acquired data typically require more iterations to converge, reaching values below, Cost $\approx -7.5$, where the ideal value is -8.5, as determined by the offset term in the Hamiltonian function. In Figure~\ref{fig:res}, we show the sampled bitstring statistics and the reconstructed density matrices for the optimal circuit depth of 3 and circuit architecture $\{\mathrm{R_z},\mathrm{R_y},\mathrm{R_z}\}$: bar plots show sampled bitstring probability weightings (with inset panels indicating the associated measurements), and panels (a)(i–ii) and (b)(i–ii) compare simulated and experimental bitstring distributions for the anti-correlated and correlated targets, respectively. In each case, the  guessed density matrix is formed from the normalised weighted sums of the reshaped bitstrings. We then reconstruct the density matrix by $\rho=\frac{\mathrm{P}^{\dagger}\mathrm{P}}{Tr(\mathrm{P}^{\dagger}\mathrm{P})}$ \cite{agnew2011tomography}, here $\mathrm{P}$ is the matrix consisting of guessed elements. This ensures physical density matrices with positive eigenvalues.
\begin{figure*}[ht!]
	\centering
\includegraphics[width=1\linewidth]{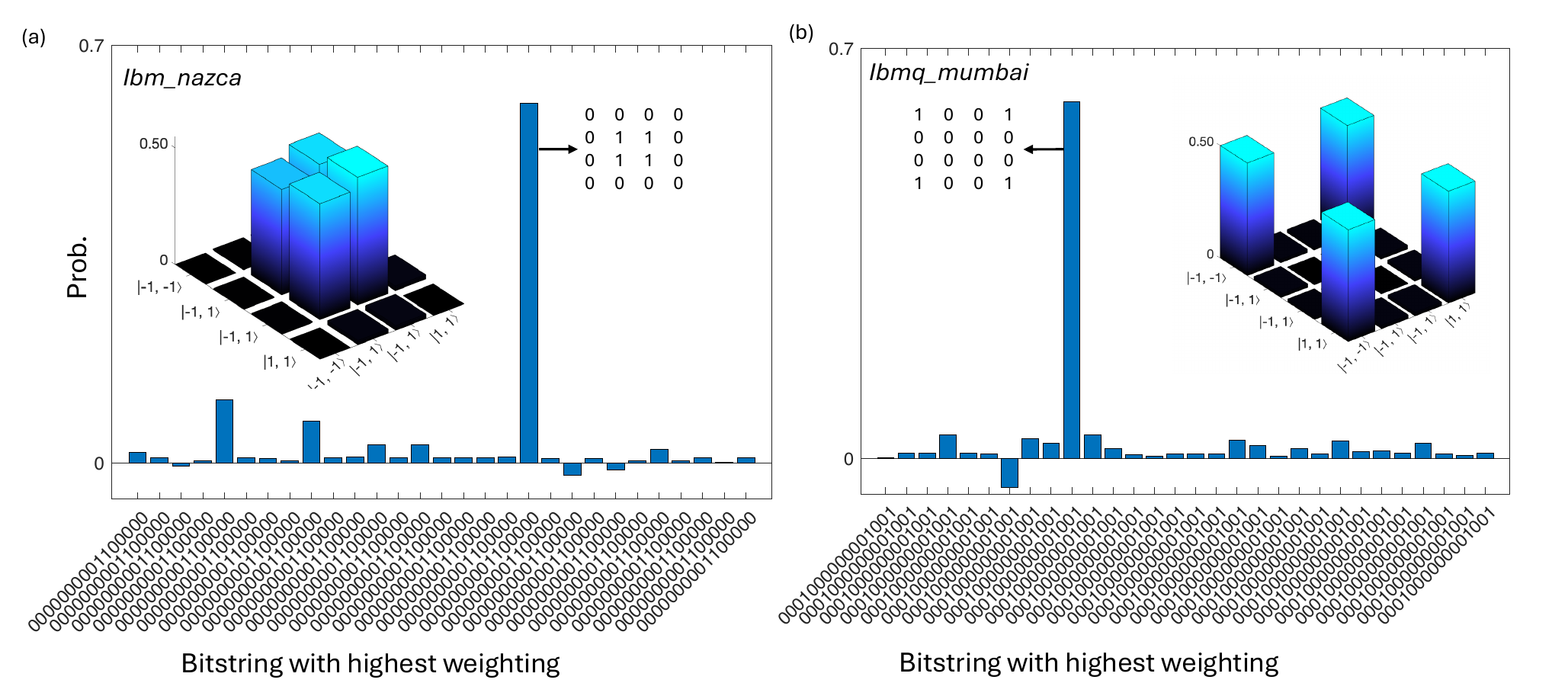}
	\caption{ {Quantum processor reconstructions: Measured bitstring distributions from (a) $ibmq\_ nazca$ and (b) $ibmq\_ mumbai$ and with level$-$1 error mitigation. The insets show the real parts of the density matrices, which are taken from the weighted sums of the reshaped binary strings.}}
		
	\label{fig:res2}
\end{figure*}
We assessed the quality of our reconstructions using the \emph{state fidelity}, defined as
\begin{equation}
F(\rho, \sigma) = \left( \mathrm{Tr}\left[\sqrt{\sqrt{\rho} \, \sigma \, \sqrt{\rho}} \right] \right)^2 ,
\end{equation}
where $\rho$ and $\sigma$ are the density matrices being compared. Fidelity ranges from $0$ (for orthogonal states) to $1$ (for identical states). In our analysis, this metric was used to benchmark the reconstructed density matrices against their respective references. For simulated input data, the reference matrices were those originally used to generate the measurement outcomes, whereas for experimental data, the reconstructions were compared against density matrices obtained via the standard maximum likelihood estimation (MLE) method.

Our VQE solutions, which were initially optimised using simulators, reliably achieved fidelities, $\approx$ 0.995, 0.999, 0.987,0.967, all of which were above 95$\%$. Next, we implemented the scheme on IBM quantum processors, namely, 
$ibmq\_mumbai$ and $ibmq \_ nazca$ {(both are now retired; we provide a .mat file containing raw data and list the hardware calibration properties in the Appendix section)}, with error-mitigation resilience level 1.The sampled bistrings and reconstructed density matrix from the ensemble of solutions are shown in Fig. \ref{fig:res2} having fidelities 0.996 and 0.995 in Fig.\ref{fig:res2} (a) and Fig.\ref{fig:res2} (b), respectively. The fidelity was computed with respect to the state determined via the traditional MLE approach (shown as insets in the same figure).


\section{Discussion}
The present study demonstrates that Quantum State Tomography (QST) can be reformulated as an Ising Hamiltonian and addressed within a variational eigensolver framework. Our demonstrations on OAM-entangled photon pairs confirm that the mapping is algebraically consistent and operational on both ideal simulations and noisy optical data. In particular, shallow, hardware-efficient single-qubit rotation architectures (notably a single $\{\mathrm{R_y}\}$ layer without entanglement) yield the fastest reconstructions from experimental coincidence counts. This supports the idea that added circuit depth or nested rotation blocks, while beneficial in noiseless simulations, become vulnerable to noise and optimization instabilities in practice.

It is important to note, however, that the present method does not yet perform full quantum state reconstruction in the usual tomographic sense. In principle, our method discretizes, or effectively binarizes, the density matrix by encoding its independent parameters into binary spin variables. Each component of the matrix elements is mapped onto a finite bitstring representation, so that the reconstruction problem can be cast as a combinatorial optimisation over these spin degrees of freedom. The resulting formulation allows the cost function to be embedded into an Ising Hamiltonian, where each spin configuration corresponds to a particular binary approximation of the density matrix. In this way, the ground-state solution identifies the bitstring that best represents the reconstructed quantum state within the resolution of the encoding. For this demonstration, the reconstructed density matrix is obtained by taking the average over all bitstring configurations, weighted by their corresponding eigenvalues of the Ising Hamiltonian. This ensures that the dominant, lowest-energy configurations contribute most strongly to the final outcome, while higher-energy (less optimal) solutions have proportionally smaller impact. In this sense, the method naturally emphasised the principal components of the reconstruction. Importantly, this reconstruction procedure does not assume that the underlying quantum state is pure. Since the Ising Hamiltonian is constructed from a Pauli decomposition of the density matrix, which is formally general, both pure and mixed states are admissible at the level of the cost function. The restriction to real-valued reconstructions in the present study arises solely from the chosen encoding of the density-matrix elements onto real binary variables, rather than from the variational framework or cost function itself. Mixed states are therefore representable within the present framework; for example, uniform or near-uniform bitstring configurations correspond, upon reshaping, to highly mixed density matrices. The primary limitation is not purity, but the finite resolution imposed by the binary encoding of the matrix elements. To systematically improve the precision of the recovered state, additional qubits can be introduced to increase the bit depth of the encoding, thereby refining the resolution with which the density matrix elements are represented

At the same time, our implementation has clear limitations. In the present implementation, the reconstruction retains only the real components of the density-matrix vectorization. This limitation originates from the real-valued binary encoding adopted in the Ising mapping, rather than from the underlying least-squares formulation, which is formally general. In principle, this limitation can be lifted by enlarging the measurement operator set to include observables whose expectation values depend explicitly on the imaginary components, or by decomposing the density matrix into real and imaginary parts and treating them as independent variables, at the expense of an increased optimization dimension. Both extensions would require a reformulation of the Hamiltonian and an increased number of encoded variables, and are therefore beyond the scope of the present proof-of-concept demonstration.

Moreover, the small two-qubit scope precludes any claim of quantum advantage over classical least-squares or maximum-likelihood approaches.  The main contribution of this work is therefore methodological, we provide a general algebraic route from projector measurements to an Ising objective that can be optimized with hybrid quantum–classical resources, without reducing the information theoretic requirements of quantum state tomography. Looking ahead, this mapping offers a flexible platform for investigating scalable encodings, ansatz design, and integration with error-mitigation strategies. As system sizes grow, such variational approaches may offer a complementary route in regimes where classical reconstruction is challenged by the high-dimensional optimization associated with the quadratic parameter scaling of general tomography. Preliminary exploratory tests on three-level (qutrit) states indicate that the approach remains operational beyond qubits, although increased noise and control errors currently limit reconstruction fidelity.  Any potential utility would arise from practical considerations such as discrete encodings or hardware native expectation value estimation rather than from guaranteed asymptotic speedups. In quantum structured light, there is already growing interest in certifying entanglement in high-dimensional systems\cite{agnew2011tomography,bavaresco2018measurements}, which highlights the need to explore alternative schemes capable of efficiently characterizing such states.

\section{Conclusion}
In summary, we have demonstrated that Quantum State Tomography (QST) can be reformulated as an Ising optimisation problem and solved within a variational eigensolver framework, with proof-of-principle demonstrations on structured photons, particularly OAM-entangled photons of different symmetries. Although our implementation is currently limited to small, real-valued reconstructions, the method provides a general algebraic route from tomographic data to Ising objectives that are optimizable using hybrid quantum–classical platforms. This establishes a foundation for exploring state reconstruction, which can be crucial for high-dimensional quantum systems that use structured light as a resource, where providing efficient state reconstruction and verification remains a pressing challenge.

\section{acknowledgments}
The authors would like to acknowledge the IBM team: Hamed Mohammadbagherpoor, Francisco Martin Fernandez, Johannes Greiner and Voica Radescu for their support and guidance on utilising the Matlab quantum computing platform. The authors also acknowledge funding from SA QuTI and the National Research Foundation (South Africa)

\section*{Data Availability Statement}
The data that support the findings of this study is available from the corresponding author upon reasonable request and the scripts used are available from the repository  \href{https://github.com/Mwezi-Koni5/Towards-reconstructing-quantum-structured-light-on-qc.git
}{Towards-reconstructing-quantum-structured-light-on-qc.git}\cite{Supplement}
\section*{Appendix}
\subsection{Hardware calibration properties.}

\begin{figure}[h!]
	\centering
	\includegraphics[width=1\linewidth]{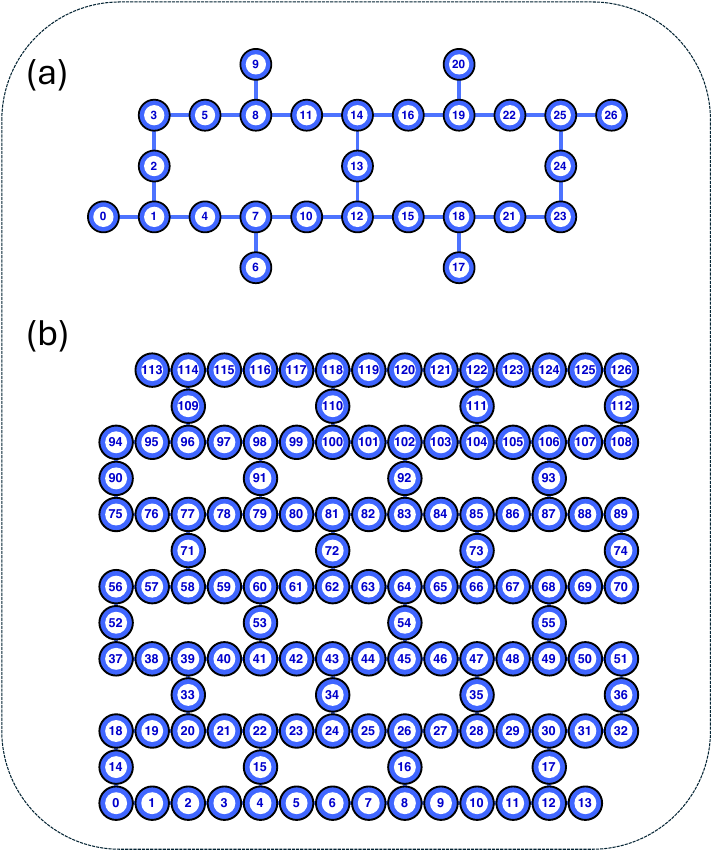}
	\caption{Qubit layouts and coupling maps of the IBM Quantum Falcon-family processors used in the demonstrations. 
    (a) \textit{ibmq\_mumbai}: 27 qubits arranged in a heavy-hexagonal lattice. 
    (b) \textit{ibmq\_nazca}: 127 qubits extending the heavy-hexagonal architecture with the corresponding coupling map. 
    Qubits are numbered for reference, and edges indicate native two-qubit interactions. Both devices are now retired; layouts are shown for reproducibility purposes.}
	\label{fig:qubit_layout}
\end{figure}

For reproducibility, we provide the key calibration metrics and connectivity layouts of the IBM Quantum Falcon-family processors used in our demonstrations. The \textit{ibmq\_mumbai} device, now retired, consisted of 27 superconducting transmon qubits arranged in a heavy-hexagonal lattice, whereas the larger \textit{ibmq\_nazca} processor, also retired, extended this architecture to 127 qubits with improved coherence times and gate fidelities. Tables~\ref{tab:ibmq_mumbai_cal} and~\ref{tab:ibmq_nazca_cal} report representative hardware parameters obtained at the time of the demonstrations, and Figure~\ref{fig:qubit_layout} illustrates their respective qubit layouts and coupling maps. These calibration metrics provide context for the computational resources and noise characteristics underlying the results presented in the main text.

\begin{table}[htb]
\centering
\caption{Summary of hardware calibration metrics for the \textit{ibmq\_mumbai} processor at the time of the demonstration. The device is a 27-qubit Falcon r5 superconducting transmon processor operated by IBM Quantum.}
\label{tab:ibmq_mumbai_cal}
\begin{tabular}{lcc}
\hline\hline
\textbf{Parameter} & \textbf{Value} & \textbf{Units} \\
\hline
Number of qubits & 27 & --- \\
Average $T_{1}$ & $9.9\times10^{1}$ & $\mu$s \\
Average $T_{2}$ & $1.53\times10^{2}$ & $\mu$s \\
Average frequency & $4.88$ & GHz \\
Average readout error & $2.8\times10^{-2}$ & --- \\
Average single-qubit gate error & $2.7\times10^{-4}$ & --- \\
Average two-qubit (CX) gate error & $4.4\times10^{-2}$ & --- \\
\hline\hline
\end{tabular}
\end{table}

\begin{table}[htb]
\centering
\caption{Summary of hardware calibration metrics for the \textit{ibmq\_nazca} processor at the time of the demonstration. The device is a 127-qubit Falcon-family transmon processor operated by IBM Quantum.}
\label{tab:ibmq_nazca_cal}
\begin{tabular}{lcc}
\hline\hline
\textbf{Parameter} & \textbf{Value} & \textbf{Units} \\
\hline
Number of qubits & 127 & --- \\
Average $T_{1}$ & $1.94\times10^{2}$ & $\mu$s \\
Average $T_{2}$ & $1.30\times10^{2}$ & $\mu$s \\
Average frequency & $5.08$ & GHz \\
Average readout error & $4.5\times10^{-2}$ & --- \\
Average single-qubit gate error & $1.3\times10^{-3}$ & --- \\
Average two-qubit (ECR) gate error & $6.3\times10^{-2}$ & --- \\
\hline\hline
\end{tabular}
\end{table}

\end{document}